\documentclass[pra,aps,twocolumn]{revtex4}
\usepackage{graphicx}
\usepackage{amsfonts}
\usepackage{amsmath}
\usepackage{dcolumn}
\usepackage{bm}
\usepackage{epsfig}
\begin{document}

\title{Negative and positive hysteresis in double-cavity optical bistability in three-level atom }
 
\author{H. Aswath Babu and  Harshawardhan Wanare}

\email[]{hwanare@iitk.ac.in}

\affiliation{Department of Physics \\ Indian Institute of Technology, Kanpur 208 016, India}

\date{\today}

\begin{abstract}
We present novel hysteretic behaviour of a three-level ladder atomic system 
exhibiting  double-cavity optical bistability in the mean-field
limit. The two fields coupling the atomic system experience feedback 
via two independent, unidirectional, single mode ring cavities and exhibit cooperative phenomena,
simultaneously.  The system displays a range of rich dynamical features varying from 
normal switching to self pulsing and a period-doubling route to chaos for both the fields.
We focus our attention to a new hump like feature in the bistable curve arising purely due to 
cavity induced inversion, which eventually leads to negative hysteresis in the
bistable response.  
This is probably the only all-optical bistable system that exhibits 
positive as well as negative bistable hysteresis in different input field intensity regimes.  
For both the fields, the switching times, the associated 
critical slowing down,  the self-pulsing characteristics, and the chaotic  behaviour can be controlled
to a fair degree, moreover, all these
effects occur at low input light levels.
\end{abstract}

\maketitle
All-optical bistability has remained a focus of research for more than four decades~\cite{szoke,gibbs,lugiato1984,joshi_modern} and apart from 
its potential application as a switch~\cite{gibbs_book} in optical communication technology it has
continued to remain a test bed for fundamental research related to cooperative
phenomena~\cite{bonifacio,vengalattore} as well as to the study of nonlinear dynamical aspects such as self-pulsing, instabilities and chaos~\cite{gibbs1981,ikeda1,ikeda2,taki}.
A deeper understanding of issues related to quantum aspects like entanglement
and cooperative  behavior in presence or absence of 
instabilities would be critical in realizing a functional quantum computer. 
Due to recent developments related to cold-atoms in optical lattices~\cite{zoller1} and 
atomic chips~\cite{folman}, the aspects related to  cooperative phenomena have become 
vital~\cite{pritchard}. Design of smaller trap size ($< \lambda$) demands a treatment that 
allows for cooperative effects, and  possibly at multiple frequencies.
Cooperative effects at multiple frequencies can be realized in multi-level atoms,
and we show the onset of instabilities in such systems even at low light levels.
Multilevel atoms have been recently used to
create and  control cooperative effects, such as in  
the multiparticle dark states~\cite{schempp} 
and Rydberg blockade effects~\cite{pfau} in cold atoms in a trap. 
In order to understand interplay of multi-colored cooperative effects
we explore the semi-classical dynamics
of three-level atoms interacting with two 
fields coupling two adjacent transitions at which they exhibit cooperative behavior simultaneously.
The atomic level structure itself provides the  coupling between the two
distinct cooperative branches.     

In an early work, one of the authors (HW with G.S. Agarwal) had shown 
the possibility of
control of optical bistability~\cite{harsha_ob} in three-level atomic medium. 
In that configuration, the optical bistability exhibited by one (probe) field 
is controlled by another (control) field coupling an adjacent transition. The field
exhibiting bistability  experiences 
conventional cavity feedback, whereas the
control field is held constant without feedback.
In order for the probe field to exhibit optical bistability (OB) 
it not only needs to interact strongly with 
the active media but it must experience sufficient
feedback (thus satisfying the conditions necessary to exhibit cooperative behavior), 
which is obtained  through  an external unidirectional ring-cavity.
Such configuration allows effective engineering of various characteristics of
OB, including tailoring of thresholds, changing On/Off intensities of the 
output field and obtaining multistability~\cite{joshi}. These effects were also experimentally
realized~\cite{xiao_ob}. Furthermore, three-level atom with a control field (without feedback)
has been shown to exhibit instability in the context of OB~\cite{yang}.

\begin{figure}[h]
\begin{center}
\includegraphics[width=0.45\textwidth]{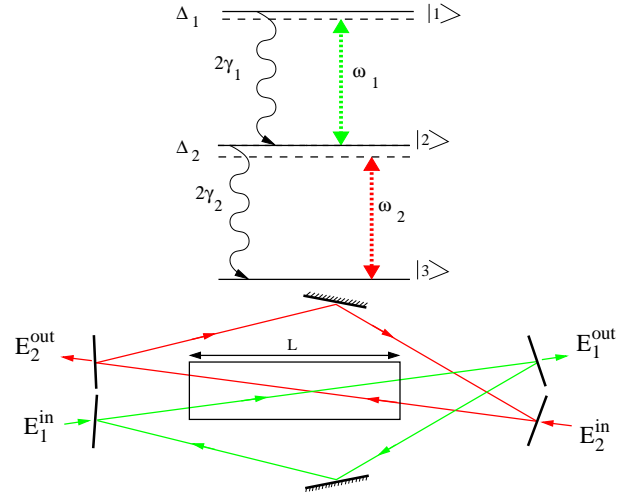}
\end{center}
\caption{(Color online) (Top) Schematic of the three-level ladder system 
interacting with the two fields $E_1$ and $E_2$ at frequencies $\omega_1$ and $\omega_2$, respectively.
(Bottom) The OB set up with the collection of atoms (within the length $L$) that interact with the two fields
in two independent unidirectional, single mode  ring cavities.}
\label{schatomfig}
\end{figure}
In this work, we again have two fields that couple two adjacent transitions in a three-level atom, however, both the fields experience feedback via two independent, single mode, unidirectional ring cavities, as indicated in Fig.~\ref{schatomfig}. This system provides independent control over the two cooperative parameters. The fields are chosen to be counter-propagating within the active medium in order to minimize the two-photon Doppler broadening in the ladder-system, however our calculation is undertaken for a homogeneously broadened atomic gas. Such a configuration leads to {\em negative} hysteresis apart from the conventional positive bistable hysteresis. As mentioned above this {\em single mode} two cavity system leads to  a variety of nonlinear dynamical effects including self-pulsing of the two fields at their distinct frequencies and  a period-doubling route to chaos in the lower cooperative branch.

Nonlinear dynamics related to OB has traditionally been associated with inclusion of multi-modal treatment of the
cavity field~\cite{lugiato1984,segard}, here, we have just two modes corresponding to the two monochromatic fields coupling the adjacent transitions. The theory of two-photon amplifiers and absorbers forms the traditional basis 
of studying these systems, which traditionally reduce  the problem to an effective two-level model and has been extensively studied~\cite{narducci,galatola}. An early work describes the possibility of obtaining chaos for atoms interacting with a single mode, where chaos occurs in the upper-branch for extremely large cooperative parameters accompanied by large atomic and cavity detunings~\cite{lugiato-chaos}. There have been other studies relating to the dispersive regime of two-photon OB which deals with the effects of cavity detuning that controls both the fields simultaneously~\cite{grangier}. In our model, the system exhibits chaos at sufficiently low-light levels in the
lower branch of the OB response without being restricted to  any special regime. Moreover, the independent cavities allow for an effective control of the feedback of the fields, thus tailoring the cooperative character. It should be noted that all the rich nonlinear dynamical features arise purely due to the {\em interplay} of the two cooperative branches.

Apart from a plethora of nonlinear dynamical features, the fields exhibit cavity induced inversion and positive as well as negative hysteresis OB, as one varies the input intensity. To the best of our knowledge, we are not aware of any other {\em all-optical} bistable system that exhibits chaos at low-light levels or even the negative and positive
hysteresis OB. Our results are also in contravention with those reported earlier, in particular we observe enhanced output field resulting from creation of population inversion between the states $|1\rangle$ and $|2\rangle$, in presence of concurrent feedback for fields coupling  {\em both} the transitions. The effects are particularly significant  as they occur at low light levels and thus could be used as a nonlinear component in optical circuitry.
The details of the nonlinear dynamical studies are presented in detail in the companion paper~\cite{aswath_2}. The atom-field density matrix equations
and the field equations governing the dynamics of the OB system in the mean-field limit are given as 
\begin{eqnarray}
\frac{\partial\rho}{\partial t} &=&-\frac{i}{\hbar} \left[{\hat H},\rho \right]+\cal{\hat L}
\rho,\nonumber\\
\frac{\partial x_{1}}{\partial  t}& = & \kappa_{1}\left[-x_{1}(1+i
\theta_{1})+y_{1} + 2 i C_{1} \rho_{12},\right] ~\label{finaleqn1}\nonumber\\
\frac{\partial x_{2}}{\partial  t}& = & \kappa_{2}\left[-x_{2}(1+i
\theta_{2})+y_{2} +  2 i C_{2} \rho_{23}.\right] 
~\label{diffeqn}
\end{eqnarray}
These equations describe the three-level atom coupled to the two fields at frequencies $\omega_1$ and $\omega_2$, which experience feedback through two independent cavities characterized by the cavity decay $\kappa_1$ and $\kappa_2$, with cavity detuning $\delta_1$ and $\delta_2$ (scaled as $\theta_i=\delta_i/\kappa_i$ where $i=1,2$) respectively. The total Hamiltonian in the dipole approximation after undertaking the rotating wave approximation
is given as: $\hat H = \hbar(\Delta_1 + \Delta_2) |1\rangle \langle 1| + \hbar  \Delta_2 |2\rangle \langle 2| -(d_{12}.E_1 |1\rangle \langle 2| + d_{23}.E_2 |2\rangle \langle 3| + \rm{h.c.})$, where $\Delta_1$ and $\Delta_2$ are the atomic detunings. The detailed density matrix equations and the scaled variables along the lines of Ref.~\cite{lugiato1984} are all explicitly given in Ref.~\cite{aswath_2}. 

We consider the mean-field limit wherein a single pass through the ring cavity only marginally affects the fields, and the strong cooperative nature arises due to the extremely large photon lifetime of the fields within the cavity, as the transmission coefficient of the cavity mirrors is chosen to be negligibly small. The last two equations in Eq.~(\ref{diffeqn}) arise due to the cavity feedback and the fields are determined self-consistently by the coupling with the three-level atom via the cooperative parameters ($C_1$ and $C_2$). Under steady state conditions one can obtain the various domains of stability for different input field strengths, we describe in detail such a stability domain map in Ref.~\cite{aswath_2}, wherein the regions of stable switching, unstable regions that exhibit  self-pulsing  and chaotic dynamics are clearly identified. The bistable behavior can be obtained for different combinations of the  input/output fields while either keeping the other input field constant ({\em i.e.} OB response of $|E_1^{in}|$ versus  $|E_1^{out}|$ as well as  $|E_2^{out}|$, while holding  $|E_2^{in}|$ constant and other such combinations)
or, varying the other input field  with a specific functional dependence. 

\begin{figure}[thb]
\includegraphics[width=0.5\textwidth]{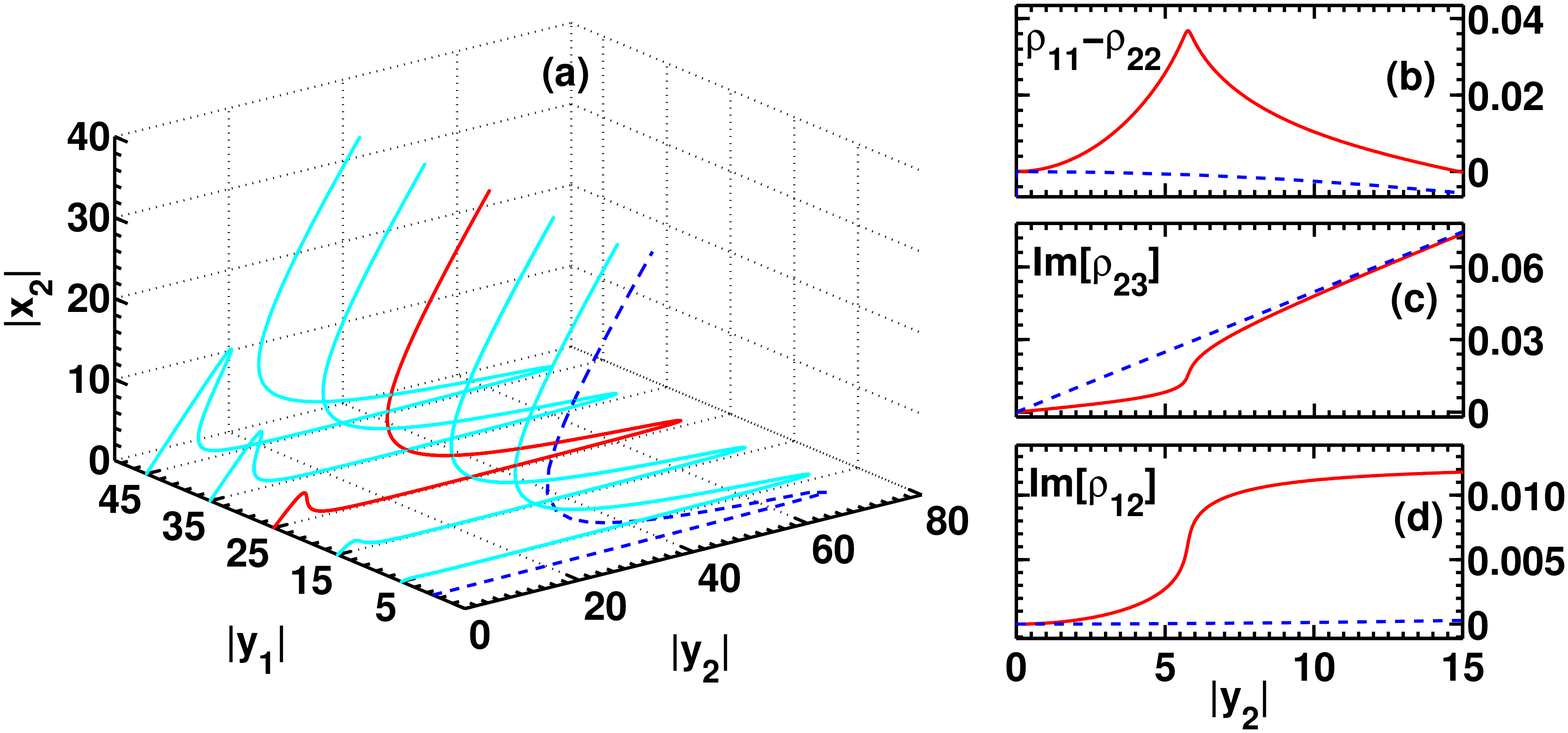}
\caption{(Color online) (a) The OB response for the $\omega_2$ field for 
various values of the input field at $\omega_1$. (b) Enhancement of the population inversion on the   $|1\rangle \leftrightarrow |2\rangle$ transition with (red)  and without (blue)  the feedback for the field at $\omega_1$. Similarly, (c) and (d) indicate the absorption $Im\{\rho_{23}\}$ and $Im\{\rho_{12}\}$, corresponding parameter values  $|y_1|=25$ and $C_1 =1000$ (red curve), for blue dashed curve $|y_1| = 0.5$ and $C_1 =0$. The other relevant parameters are $C_1=1000,C_2=100,\kappa_i=1,\gamma_i=1,\Delta_i=0,\theta_i=0$ for $i=1,2$}
\label{enha} 
\end{figure}

The output levels, the threshold of switching, the range of input fields exhibiting bistability are all dependent on the parameters like the decay rates (both atomic and the cavity), the detunings (both atomic and the cavity) and the cooperative parameters corresponding to the two transitions. The novel effects arise due to the inclusion of feedback for both the fields that couple the three-level atoms. The two fields being complex (on interaction) their relative phases and strengths become critical factors in determining the dynamics. In the context of nonlinear dynamics the corresponding phase space, as well as, the parameter space is exceedingly large owing to fourteen different physical parameters ($\gamma_{1,2}$, $\Delta_{1,2}$, $\theta_{1,2}$, $\kappa_{1,2}$, $C_{1,2}$,  $E^{in}_{1,2}$) all of which can be varied independently. 
Unlike in the OB with a control field (without feedback, as in Ref.~\cite{harsha_ob}) it is the simultaneous interaction of the atom with both the fields and their {\em independent} feedback from the cavities that self-consistently determine the output fields (both their phase and amplitude). Hence, with regard to the numerical implementation we express a note of caution as this setup does not permit an {\em apriori} choice of both the (complex) output fields. In the conventional computation of OB (in steady state and in the mean field limit) involving feedback for one field, one usually specifies the output field (which could be chosen to be a real value) and calculate the requisite unique (complex) input field. Such strategy is conventionally adopted due to the multivalued nature of the output field, because of the input-output relationship in form of the $S-$ shaped OB curve; however a given output field uniquely determines the input field. A generalization of a such a strategy fails 
because it is impossible to choose the amplitude and phase of both the output fields apriori. The input fields, the cavity fields and their interaction with the medium self-consistently determine the amplitudes and phases of output fields. Without taking into account the complex nature of the fields within the cavity one would miss out the hump like feature (discussed below) at low light levels~\cite{xiao-joshi} which is crucial and eventually transforms into the negative hysteresis as shown in Fig.~\ref{enha}(a). In order to deal with such numerical constraints we used the 
Newton-Raphson method to obtain the steady state solution of the nonlinear atom-field equations along with the boundary conditions. As expected an {\em arbitrary} choice of relative strength and phases of the output field does not necessarily correspond to a physically viable input field variation.

In order to mimic a typical experimental situation we use real values for the input fields and compute the resultant complex  output fields self-consistently along the different branches of the S-shaped OB response. We also note that in order to focus our attention to the new low input light OB regime in this paper, we have avoided the regimes involving multi-stability which is easily obtained in this double cavity OB system. There are two important features that we highlight here. The first one is the enhancement of the output field at $\omega_2$ coupling the $|2\rangle \leftrightarrow |3\rangle $ transition arising due to the creation of inversion in the $|1\rangle \leftrightarrow |2\rangle $  transition which leads to the hump like response and eventually to negative hysteresis OB. The second feature is the novel switching characteristics wherein the output fields exhibit self-pulsing, moreover the two fields completely mimic each other in their temporal response. The hump like feature in the OB response indicates an enhancement of the field at $\omega_2$ arising due to suppressed absorption along the $|2\rangle \leftrightarrow |3\rangle $ transition. This  occurs in the lower cooperative branch at low intensities of the $\omega_2$ input field, where initially the population is dominantly in the ground state $|3\rangle$. A large cooperative parameter $C_1$ along the upper transition $|1\rangle \leftrightarrow |2\rangle$ leads to enhanced interaction with the $\omega_1$ field  resulting in the extraction of a significant fraction of the population into the excited states. Furthermore, as the upper state ($|1\rangle $) population builds up there is a lowering of the influence of the field at $\omega_2$ on these atoms. This leads to the creation of inversion along the upper transition $|1\rangle \leftrightarrow |2\rangle $ {\em i.e. $\rho_{11} > \rho_{22}$}. This dynamics occurs due to the asymmetric choice of the cooperative parameters $C_1 > C_2$ at the cavity resonant condition $\theta_1=\theta_2 = 0$. However, similar dynamics can be obtained for comparable values of $C_1$ and $C_2$ in the bad cavity limit, {\em i.e.} with finite cavity detuning $\theta_2$~\cite{aswath_2}. The decrease in absorption ($Im\{\rho_{23}\}$) of the $\omega_2$ field, accompanied by enhanced absorption ($Im\{\rho_{12}\}$) at the $\omega_1$ field  {\em  with} and {\em without} the feedback for the $\omega_1$ field clearly demonstrates the reliance of the enhancement of the field at $\omega_2$ on 
cavity assisted inversion [see Figs.~\ref{enha}(b),(c) and (d)]. With increasing incident field $y_1$, the above effects are enhanced and the hump like feature becomes quite exaggerated and eventually results in OB with a negative-hysteresis loop as indicated in Fig.~\ref{enha}(a). A similar scenario is observed with increasing cooperative parameter $C_1$ of the upper transition $|1\rangle \leftrightarrow |2\rangle$. 

\begin{figure}[thb]
\includegraphics[width=0.5\textwidth]{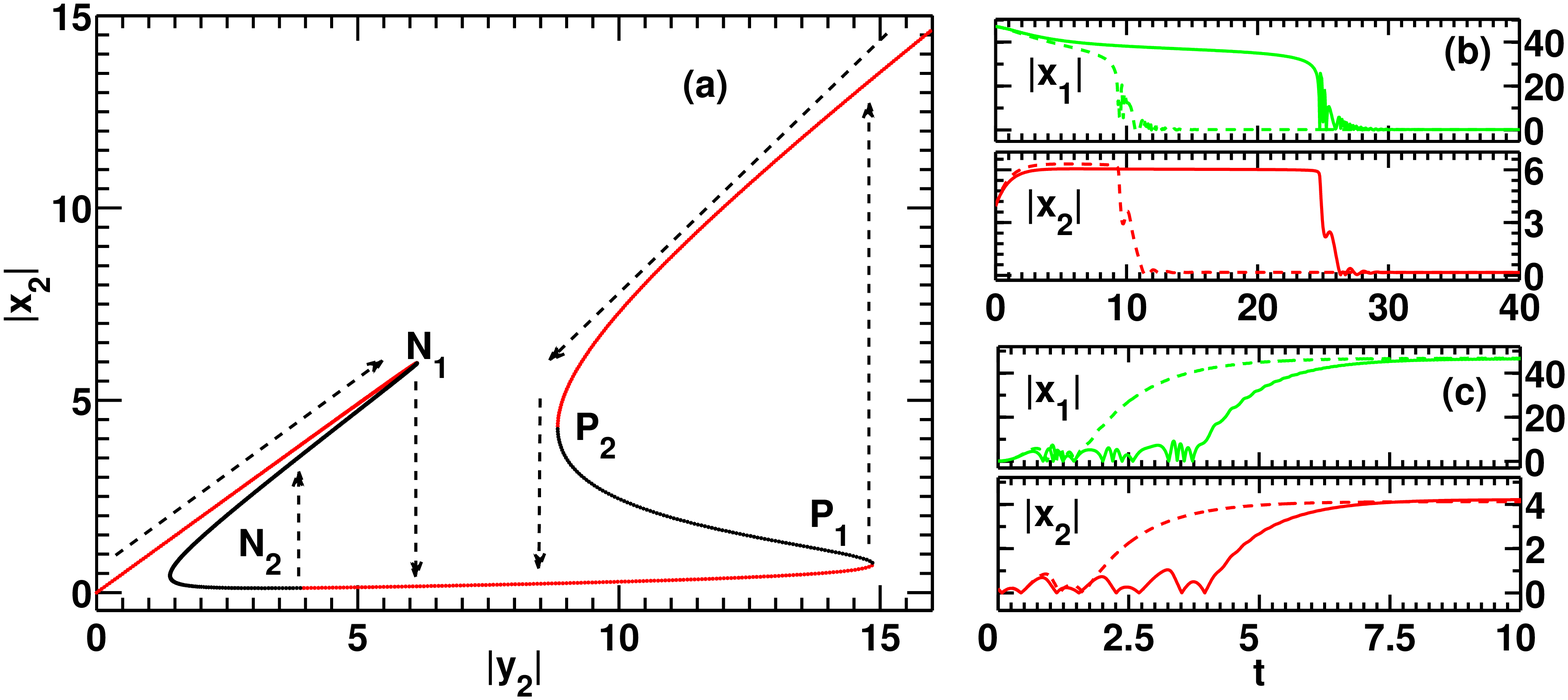}
\caption{(Color online) (a) The OB response for the $\omega_2$ field which exhibits 
negative as well positive hysteresis (for $C_1=5000,C_2=20,|y_1|=50$).  The color red (black) indicate stable(unstable) steady state response, and the arrows indicate the associated switching. (b) and (c) Critical slowing down associated with the negative hysteresis along  $N_1$ and $N_2$, respectively. The time evolution of fields are indicated by dashed lines for the operating points $|y_{2}|_{op} = 6.5(N_1)$ and $3.5(N_2)$ and the solid lines indicate operating points $|y_{2}|_{op} = 6.2(N_1)$ and $3.6(N_2)$. The corresponding threshold points are $|y_{2}|_{th} = 6.13(N_1)$ and $4.05(N_2)$. The other relevant parameters are same as in Fig.~\ref{enha}}
\label{crit} 
\end{figure}

We describe in detail the negative hysteresis loop arising in this system (Fig.~\ref{crit}). As described above the field at $\omega_2$ is enhanced, however, on further increase in the input field $y_2$ the population from the upper states is drawn back into the lower levels $|2\rangle$ and $|3\rangle$ ultimately leading to large absorption of the $\omega_2$  field, and the output switches to the {\em off-} state (indicated by $N_1$ in Fig.~\ref{crit}(a)). In the reverse direction as the input intensity $y_2$ is decreased the output field switches from an {\em off-} state to an {\em on-} state  along $N_2$ (different from $N_1$) due to the large $\omega_1$ field already circulating in the other cavity, thus encompassing within it a negative-hysteresis loop. This hysteresis is exactly opposite to the usual bistability (we denote as the  positive hysteresis) wherein low input intensities leads to low output intensity and only for larger input intensities the transition saturates (for zero atomic detuning) leading to the large output field and the corresponding reverse loop that encloses a hysteresis (indicated as $P_1$ and $P_2$ in Fig.~\ref{crit}(a)). This conventional positive hysteresis OB occurs at higher input field intensities where larger input field results in large output field and vice-versa. 

The nature of threshold points associated with the negative hysteresis are similar to the positive hysteresis of the conventional OB and displays critical slowing down. The study of the time-dependent switching indicates that, as one chooses the operating point $|y_2|_{op}$ closer to the threshold of switching one observes an increase in the time required to switch to the steady state [Fig.~\ref{crit}(b-c)]. Note that this behavior occurs at all the four threshold points $|y_2|_{th}$ associated with the switching transitions $N_{1,2}$ and $P_{1,2}$. The variation of any other parameter results in shifting of the threshold point itself and thus the associated change in the switching times due to critical slowing down. 

Apart from such multicolored stable switching there is a wealth of dynamics this system can display. It exhibits periodic self-pulsing in the steady state for both the fields. A constant input intensity at both the frequencies $\omega_1$ and $\omega_2$ results in periodic output that largely mimics each other, and the periodicity can be controlled using the cavity parameters (such as $\kappa_{1,2}$ and $\theta_{1,2}$). These detailed nonlinear dynamical aspects are discussed in the companion paper~\cite{aswath_2}. Before we conclude, we would like to point out that there is nothing special about the two-photon resonant condition that we have used in the calculations presented here, similar results are obtained even in the non-resonant cases. The nonlinear dynamics can be obtained
in a robust manner in a wide variety of regimes with appropriately chosen parameters.

In conclusion, we have demonstrated a simple all optical double cavity OB system that exhibits negative as well as positive hysteresis. We self-consistently determine the amplitude and phases of both the output fields. A novel region of response involving a hump like feature in the S-shaped OB curves, as well as negative hysteresis is demonstrated at low input light levels. These effects are a consequence of the cavity induced inversion arising from 
the simultaneous cooperative coupling at two different frequencies. The system also exhibits a range of nonlinear dynamical features such as self-pulsing and chaos, again at low input light levels.


\begin{thebibliography}{99}


\bibitem{szoke} A. Szoke, V. Daneu, J. Goldhar and N.A. Kurnit, Appl. Phys. Lett. {\bf 15}, 376 (1969).
\bibitem{gibbs} H. M. Gibbs, S. L. McCall, and T. N. C. Venkatesan
Phys. Rev. Lett. {\bf 36}, 1135 (1976).
\bibitem{lugiato1984}
L.A. Lugiato, {\it Progress in Optics}, edited by E.Wolf (North-Holland. Amsterdam) {\bf Vol.XXI}, 69(1984).

\bibitem{joshi_modern}
Joshi, Amitabh and Xiao, Min, Journal of Modern Optics. {\bf 57: 14}, 1196 (2010).
\bibitem{gibbs_book}

H. M. Gibbs, {\em Optical Bistability: Controlling Light with Light}, (Academic, New York, 1985). 
\bibitem{bonifacio} R. Bonifacio and L. A. Lugiato,
Phys. Rev. A {\bf 18}, 1129 (1978) 
\bibitem{vengalattore}
M. Vengalattore, M. Hafezi, M. D. Lukin, and M. Prentiss 
Phys. Rev. Lett. {\bf 101}, 063901 (2008). 
\bibitem{gibbs1981}
H.M. Gibbs, F.A. Hopf, D.L. Kaplan and R.L. Shoemaker,
Phys. Rev. Lett. {\bf 46}, 474 (1981). 
\bibitem{ikeda1}
K. Ikeda, H. Daido,  and O. Akimoto,
Phys. Rev. Lett. {\bf 45}, 709 (1980).  
\bibitem{ikeda2}
H. Nakatsuka, S. Asaka, H. Itoh, K. Ikeda, and M. Matsuoka,
Phys. Rev. Lett. {\bf 50}, 109 (1983).  
\bibitem{taki} M. Taki, 
Phys. Rev. E {\bf 56}, 6033 (1997).  
\bibitem{zoller1} D. Jaksch, C. Bruder, J. I. Cirac, C. W. Gardiner, and P. Zoller
Phys. Rev. Lett. {\bf 81}, 3108 (1998). 
\bibitem{folman} R. Folman, P. Kruger, D. Cassettari, B. Hessmo, T. Maier, and J. Schmiedmayer,  
Phys. Rev. Lett. {\bf 84}, 4749 (2000). 
\bibitem{pritchard} J. D. Pritchard, D. Maxwell, A. Gauguet, K. J. Weatherill, M. P. A. Jones and C. S. Adams,
arXiv:1006.4087v1 [quant-ph].
\bibitem{schempp}
H. Schempp, G. Gunter, C. S. Hofmann, C. Giese, S. D. Saliba, B. D. DePaola, T. Amthor, and M. Weidemuller,
S. Sevincli, and T. Pohl,
Phys. Rev. Lett. {\bf 104}, 173602 (2010).  
\bibitem{pfau}
R. Heidemann, U. Raitzsch, V. Bendkowsky, B. Butscher, R. Low, L. Santos, and T. Pfau,
Phys. Rev. Lett. {\bf 99}, 163601 (2007).  
\bibitem{harsha_ob}
 W. Harshawardhan and G.S. Agarwal, Phys. Rev. A, {\bf 53}, 1812 (1996).
\bibitem{joshi}
A. Joshi and M. Xiao, {\it Progress in Optics},edited by E.Wolf(North-Holland. Amsterdam) {\bf Vol. 49}, 97 (2006).
\bibitem{xiao_ob} A. Joshi,  A. Brown, H. Wang and M. Xiao,  Phys. Rev. A, {\bf 67}, 041801 (2003).
\bibitem{yang} 
 W. Yang, A. Joshi, and M. Xiao,  Phys. Rev. Lett., {\bf 95}, 093902 (2005).
\bibitem{segard} B. Segard, B. Macke, L.A. Lugiato, F. Prati and M. Brambilla,
Phys. Rev. A {\bf 39}, 703 (1989).
\bibitem{narducci} L. M. Narducci,  W. W. Eidson, P. Furcinitti
and D. C. Eteson, Phys. Rev. A {\bf 16}, 1665 (1977).
\bibitem{galatola} P. Galatola, L.A. Lugiato, M. Vadacchino and N.B. Abraham,  
Optics Commun., {\bf 69} 414 (1989).
\bibitem{lugiato-chaos}
 L.A. Lugiato, L.M. Narducci, D.K. Bandy, C.A. Pennise, Optics Commun., {\bf 43}, 281(1982).  
\bibitem{grangier} P. Grangier, J. F. Roch, J. Roger, L. A. Lugiato, E. M. Pessina, G. Scandroglio and P. Galatola,  
 Phys. Rev. A {\bf 46}, 2735 (1992).    
 
\bibitem{aswath_2} H. Aswath Babu, and Harshawardhan Wanare, arXiv:1009.1550v1 [nlin.CD] 

\bibitem{xiao-joshi}
 A. Joshi, and M. Xiao,  Appl. Phys. B, {\bf 79}, 65 (2004).

\end{thebibliography}
\end{document}